\documentclass[english]{article}
\usepackage[T1]{fontenc}
\usepackage[latin9]{inputenc}
\usepackage{amsmath}
\usepackage{amssymb}
\usepackage{stmaryrd}
\usepackage{babel}
\begin{document}
\title{General Shells and Generalized Functions}
\author{Albert Huber\thanks{hubera@technikum-wien.at}$\,$\thanks{albert.huber@uni-graz.at}}
\date{{\footnotesize UAS Technikum Wien, Höchstädtplatz 6, 1200 Vienna, Austria;
Karl-Franzens-University Graz, Schubertstraße 51, Graz, Austria}}
\maketitle
\begin{abstract}
In this work, standard methods of the mixed thin-shell foramlism are
refined using the framework of Colombeau's theory of generalized functions.
To this end, systematic use is made of smooth generalized functions,
in particular regularizations of the Heaviside step function and the
delta distribution, instead of working directly with the corresponding
Schwartz distributions. Based on this change of method, the resulting
extended thin shell formalism is shown to offer a decisive advantage
over traditional approaches to the subject: it avoids dealing with
ill-defined products of distributions in the calculation of nonlinear
curvature expressions, thereby allowing for the treatment of problems
that prove intractable with the 'conventional' thin-shell formalism.
This includes, in particular, the problem of matching singular spacetimes
with distributional metrics (containing a delta distribution term)
across a joint boundary hypersurface in spacetime, the problem of
setting up the dominant energy condition for thin shells, and the
problem of defining reasonably rigorously nonlinear distribution-valued
curvature invariants needed in higher-derivative theories of gravity.
Eventually, as a further application, close links to Penrose's cut-and-paste
method are established by proving that results of said method can
be re-derived using the generalized formalism presented.
\end{abstract}
{\footnotesize\textit{Key words: Colombeau algebras, generalized functions,
thin shell formalism, gluing methods;}}{\footnotesize\par}

\section*{Introduction}

The thin-shell formalism has proven time and again to be an indispensable
asset in many relativists' toolkits \cite{barrabes1991thin,darmois1927equations,israel1966singular,mars1993geometry,poisson2004relativist}.
This is not least due to the fact that said formalism provides powerful
methods for gluing together spacetime partitions across a joint boundary
hypersurface and, at the same time, the necessary junction conditions
for such gluing. On top of that, the formalism enables the construction
of novel weak (distributional) solutions of Einstein's gravitational
field equations, which arise naturally whenever pairs of spacetimes
cannot be glued together smoothly. All these solutions have one thing
in common: they lead to thin shells of matter concentrated on joint
portion of the boundary hypersurface (or on a series of non-overlapping
hypersurfaces) in spacetime. As such, they are typically invoked for
the physical description of infinitesimally thin concentrations of
massless particles, plane-fronted impulsive (pp) waves or other types
of confined singular gravitational fields \cite{barrabes2001detection,barrabes2002impulsive,barrabes1991thin}.

Mathematically, due to their singular nature, thin shells must be
treated within the framework of linear distribution theory. At first
glance, this seems to be bad news: after all, Einstein's theory of
gravity is a non-linear theory that requires the possiblity of multiplying
distributions at the same spacetime point. According to Schwartz's
impossibility result \cite{schwartz1954limpossibilite}, however,
the latter proves impossible, thus prompting the question of whether
non-smooth gluings of spacetimes can be meaningfully treated in Einstein-Hilbert
gravity at all. 

Yet, as is well known, the matter is not as bad as it looks at first
sight: By systematically imposing junction conditions on the first
and second fundamental forms of the spacetimes to be glued (or on
tensorial quantities closely related to these quantities), the mentioned
problems can be brought under control to a reasonable extent. This
is because the imposed conditions ensure that the correction terms
arising in the spacetime curvature and Einstein's field equations
at most proportional to the Dirac delta distribution, but not to its
derivatives. As a consequence, the method used yields mathematically
perfectly well-defined tensor distributions and associated locally
singular solutions to the field equations in the presence of a non-smooth
gluing of spacetime partitions across their joint boundary hypersurface;
provided at least that the spacetime metrics to be glued are smooth
and minimally twice differentiable. 

And even if everything thus appears to be perfectly in order with
the formalism mentioned, there are still some serious shortcomings
that should clearly be remedied, but which have not yet been addressed
in the literature. 

Among these shortcomings is, for example, the issue that matching
singular spacetimes with metrics of low regularity (such as, in particular,
$C^{1,1}$-Lorentzian metrics) still proves impossible with traditional
methods of the theory of thin shells. Furthermore, there is the problem
that nonlinear curvature expressions cannot be meaningfully calculated
with same formalism, since such calculations result in ill-defined
products of delta distributions that cannot be defined with conventional
linear distribution theory. Said products are encountered, for example,
in setting up the dominant energy condition or in calculating the
Kretschmann scalar for checking the singularity structure of spacetime.
Also, they are ubiquitous in higher-derivative theories of gravity
in which non-smooth gluings of spacetimes are considered \cite{reina2016junction,senovilla2013junction,senovilla2015double,senovilla2018equations}. 

Another shortcoming of the thin-shell formalism is that (in its present
formulation) it does not allow for the derivation of familiar solutions
of Einstein's field equations for confined matter fields that can
be obtained with closely related gluing methods. In particular, one
would expect the thin-shell formalism to be closely related to Penrose's
cut-and-paste method, which is defined in terms of a metric with a
Dirac delta distribution with support on the shell. Interesting types
of geometries that have been constructed with this method are, most
notably, gravitational shock wave spacetimes in flat Minkowski spacetime
and in various black hole and cosmological backgrounds. This includes
the geometries of Dray and 't Hooft, Sfetsos, and other more recently
derived types of gravitational shock waves in stationary axisymmetric
black hole backgrounds \cite{aichelburg1971gravitational,dray1985gravitational,groah2007shock,huber2021gravitational,penrose1972geometry,sfetsos1995gravitational,stephani2009exact}.
While the problem of deriving such solutions by thin-shell methods
is, as should be mentioned, treated in considerable detail in \cite{manzano2021null,manzano2024abstract},
it appears that, apart from connections to Penrose's original work
and non-expanding impulsive gravitational waves in cosmological backgrounds,
not too many results in that direction are known as of yet. 

To address this particular problem, as well as all other problems
described so far, the strategy followed in this work is to incorporate
Colombeau's mathematical theory of generalized functions \cite{colombeau2000new,colombeau2011elementary,grosser2001geometric,grosser2012global,grosser2001foundations,grosser2002global,kunzinger1999rigorous,kunzinger2002foundations,kunzinger2002generalized,kunzinger2003intrinsic,steinbauer2010geometric,steinbauer2006use}
into the gluing formalism used; culminating in a better tractable
matching approach that includes the conventional thin-shell formalism
as a special case. Though Colombeau algebras have previously been
used in the literature on the subject, in particular in the context
of signature-changing spacetimes \cite{mansouri1996equivalence,mansouri2000new,silva2024new},
no systematic approach seems to have yet been developed that attempts
to overcome persistent mathematical shortcomings of the formalism
by turning to algebras of generalized functions. 

To close this literature gap, the first section of this work presents
a generalized matching approach that avoids the direct use of Schwartz
distributions in joining metrics and curvature fields of local spacetime
partitions in favor of using manifold-valued generalized functions
(i.e., regularizations of the Heaviside step function and the delta
distribution) that converge to these distributions in a suitable Colombeau
limit. As only shown in section two of the article, this type of approach
allows Lorentzian metrics of low regularity to be joined and, at the
same time, to avoid ill-defined products of distributions (such as,
in particular, 'squares of the delta distribution') in the calculation
of nonlinear curvature expressions and in the dominant energy condition
for thin shells. Moreover, by drawing connections between the theory
of null shells and the so-called generalized Kerr-Schild framework
\cite{balasin2000generalized,huber2020distributional}, the approach
is shown to allow for the derivation of various cut-and-paste geometries
known from the literature \cite{aichelburg1971gravitational,dray1985gravitational,huber2021gravitational,penrose1972geometry,podolsky2002exact,podolsky2022penrose,podolsky2017penrose,podolsky1998impulsive,Samann:2023bko}.
As a bonus, it is shown that various generalizations of the mixed
thin-shell formalism can be given that, as it seems, have not yet
been considered in the literature on the subject. These generalizations
lead naturally to new types of thick shells whose supports are confined
to four-dimensional submanifolds rather than to single hypersurfaces
in spacetime. Physical applications of such thick shells are biefly
discussed towards the end of section two of the work. Further applications
are discussed in the third and final section of the article, with
focus being placed on the problem of matching gravitational shock
wave and pp-wave geometries, prior to the paper closing with a summary
of the results obtained.

\section{Colombeau Algebraus and Distributional Matching: Extended Thin Shell
Formalism}

The main objective of this section is to introduce a matching approach
based on of algebras of manifold-valued generalized functions \cite{grosser2001geometric,kunzinger2003intrinsic},
extending the mixed thin-shell formalism introduced in \cite{mars1993geometry}. 

To cover some relevant aspects of Colombeau's theory needed in the
following, consider a paracompact $C^{\infty}$-manifold $X$ and
an associated commutative, associative and unital differential algebra,
$\mathcal{G}(X)$, which contains the vector space of Schwartz distributions
as a linear subspace, and the space of smooth functions as a faithful
subalgebra. This so-called Colombeau algebra consists of one-parameter
families of $C^{\infty}$-functions $(f_{\epsilon})_{\epsilon}$,
which are subject to certain growth conditions in $\epsilon$. To
be more precise, $\mathcal{G}(X)$ results from forming the quotient
algebra $\mathcal{E}_{m}(X)/\mathcal{N}(X)$ of the algebra of nets
of moderate functions $\mathcal{E}_{m}(X)=\{(f_{\epsilon})_{\epsilon}\in C^{\infty}(X):\forall K\subset\subset X\:\forall P\in\mathcal{P}(M)\:\exists l\:\underset{x\in K}{\sup}\vert Pf_{\epsilon}\vert=O(\epsilon^{-l})\}$
by the ideal of nets of so-called negligible functions $\mathcal{N}(X)=\{(f_{\epsilon})_{\epsilon}\in C^{\infty}(X):\forall K\subset\subset X\:\forall m\:\forall P\in\mathcal{P}(M)\:\underset{x\in K}{\sup}\vert Pf_{\epsilon}\vert=O(\epsilon^{m})\}$,
where, in this context, $\mathcal{P}(X)$ denotes the space of all
linear differential operators on the manifold $X$. The resulting
algebra $\mathcal{G}(X)$ of generalized functions therefore comprises
expressions (regularizations of distributions in the standard physicist's
parlance) that are singular in a fixed, but in principle arbitrary
real parameter $\epsilon$ and are (weakly) equivalent to Schwartz
distributions in the limit $\epsilon\rightarrow0$ (if such a limit
exists). 

A key advantage of working with regularized objects compared to the
direct use of Schwartz distributions is that elements of $\mathcal{G}(X)$
are much easier to handle mathematically than those distributions
to which they are associated. In particular, said objects can be multiplied
with each other in a simple way, yielding meaningful results in the
Colombeau limit $\epsilon\rightarrow0$ in many cases of interest.
The use of these objects, however, reveals an uncommon subtlety: Depending
on the specific choice of regularization, different results can be
obtained in modelling one and the same product in the limit $\epsilon\rightarrow0$.
The main reason for this is that there are actually many possibilities
of how to use generalized functions to model products of distributions
and, accordingly, many different types of Colombeau algebras $\mathcal{G}(X)$
that can be constructed in different ways; all of which are a priori
equally meaningful from a mathematical point of view, though not necessarily
from a physical perspective. See here, for example, \cite{grosser2001geometric}
for further information. 

As far as the present work is concerned, elements of a given Colombeau
algebra $\mathcal{G}(\Omega)$ with $\Omega\subseteq\mathcal{M}$
will be used to replace the Heaviside step function $\theta(x)$ by
a generalized function $(\vartheta_{\epsilon}(x))_{\epsilon}$, which
matches said step function in the Colombeau limit $\epsilon\rightarrow0$
and whose first derivative $(\vartheta'_{\epsilon}(x))_{\epsilon}$
reduces to the delta distribution $\delta(x)$ in the same limit.
This is tantamount to requiring that

\begin{align}
\underset{\epsilon\rightarrow0}{\lim}\int\vartheta_{\epsilon}\,\nu_{\varphi} & \equiv\underset{\epsilon\rightarrow0}{\lim}\langle\vartheta_{\epsilon},\nu_{\varphi}\rangle\equiv\langle\theta,\nu_{\varphi}\rangle,\\
\underset{\epsilon\rightarrow0}{\lim}\int\vartheta'_{\epsilon}\,\nu_{\varphi} & \equiv\underset{\epsilon\rightarrow0}{\lim}\langle\vartheta'_{\epsilon},\nu_{\varphi}\rangle\equiv\langle\delta,\nu_{\varphi}\rangle\nonumber 
\end{align}
is fulfilled for all compactly supported one-densities $\nu_{\varphi}$
on $\mathcal{M}$ with $\varphi\in C_{0}^{\infty}(\mathcal{M})$,
where $\theta$ is the Heaviside step function and $\delta$ the Dirac
delta distribution. That is to say, the pairs $(\vartheta_{\epsilon})_{\epsilon}$,
$(\vartheta'_{\epsilon})_{\epsilon}$ and $\theta$, $\delta$ are
required to be associated in a distributional sense, i.e.
\begin{equation}
\underset{\epsilon\rightarrow0}{\lim}\,\vartheta_{\epsilon}\approx\theta,\;\underset{\epsilon\rightarrow0}{\lim}\,\vartheta'_{\epsilon}\approx\delta.
\end{equation}
These association relations prove valid already as regards elements
of a Colombeau algebra $\mathcal{G}(\Omega)$ that is constructed
in the following sense: Given a suitable smooth test function $\varphi(x)\in C_{0}^{\infty}(\Omega)$
with $\Omega\subset\subset\mathbb{R}^{n}$ having the properties 
\begin{align}
i)\;\int\varphi(x)dx^{n} & =1,\\
ii)\;\int\varphi(x)x^{\alpha} & dx^{n}=0,\forall\vert\alpha\vert\geq1,\nonumber 
\end{align}
so that 

\begin{equation}
\varphi_{\epsilon}(x)=\epsilon^{-m}\varphi(\frac{x}{\epsilon}),
\end{equation}
applies, the convolution 
\begin{equation}
\vartheta_{\epsilon}(x)=(\theta\ast\varphi_{\epsilon})(x)=\int\theta(y)\varphi_{\epsilon}(x-y)d^{n}y=\epsilon^{-m}\int\theta(y)\varphi(\frac{x-y}{\epsilon})d^{n}y
\end{equation}
defines a theta sequence $(\vartheta_{\epsilon})_{\epsilon}(x)$.
Assuming that expression $(4)$ constitutes a suitable mollifier,
it becomes clear that 
\begin{equation}
\delta_{\epsilon}(x)=(\delta\ast\varphi_{\epsilon})(x)=\int\delta(y)\varphi_{\epsilon}(x-y)d^{n}y=\epsilon^{-n}\int\delta(y)\varphi(\frac{x-y}{\epsilon})d^{n}y
\end{equation}
defines a strict delta net $(\delta_{\epsilon})_{\epsilon}(x)$ that
is associated in a distributional sense with the derivative $(\vartheta'_{\epsilon})_{\epsilon}(x)$
of $(5)$. Suitable examples of mollifiers that can be used for the
construction of these generalized functions are discussed e.g. in
\cite{grosser2001geometric}. As an explicit example, one could consider,
for instance, the bump function $\varphi(x)\in C_{0}^{\infty}(\mathbb{R}^{n})$
\begin{equation}
\varphi(x):=\begin{cases}
\overset{C\cdot exp\left(\frac{1}{\vert x\vert^{2}-1}\right)}{\underset{0}{}} & \overset{if\;\vert x\vert<1}{\underset{if\;\vert x\vert\geq1}{}}\end{cases},
\end{equation}
which is defined with respect to a constant $C$ that is chosen in
such a way that the normalization condition in $(3)$ is met. Using
either this or another mollifier to construct a Colombeau algebra
$\mathcal{G}(\Omega)$ with $\Omega\subset\subset\mathbb{R}^{n}$,
the system of association relations 
\begin{align}
\underset{\epsilon\rightarrow0}{\lim}\,\vartheta_{\epsilon}\cdot\vartheta_{\epsilon} & \approx\theta^{2}\approx\theta,\;\underset{\epsilon\rightarrow0}{\lim}\,\vartheta_{\epsilon}\cdot\vartheta'_{\epsilon}\approx\theta\cdot\delta\approx A\cdot\delta,\\
\underset{\epsilon\rightarrow0}{\lim}\,\vartheta'_{\epsilon}\cdot\vartheta' & _{\epsilon}\approx\delta^{2}\approx B\cdot\delta,\nonumber 
\end{align}
can be set up, the validity of which will prove to be absolutely decisive
in the following. The concrete values of the constants $A$ and $B$
occurring in these relations, as may be noted, depend on the choice
of the mollifier used (and thus on the choice of regularization),
whereby it may very well occur that $B=0$ applies in selected special
cases. The latter, however, is not required in the following.

As may further be noted in this context, the considered association
relations can, of course, also be set up in analogous form in the
framework of manifold-valued generalized functions \cite{kunzinger2009sheaves,kunzinger2003intrinsic}
using objects of the form $(1)$. The fact that the concrete forms
of the results, i.e. the concrete values of $A$ and $B$, strongly
depend on the type of modeling, i.e. the choice of mollifier, can
readily be substantiated by calculating the generalized expression
$\theta\cdot\delta$ , the latter of which shall now be picked out
as an example. As a basis for this, let it first be noted that

\begin{align}
\theta^{n} & \thickapprox\theta,\\
n\theta^{n-1} & \theta'\thickapprox\theta',\nonumber 
\end{align}
applies in a distributional sense. Yet, after using the fact that
$\theta'\thickapprox\delta$, this leads to the different results

\begin{equation}
\theta\cdot\theta'\thickapprox\frac{1}{2}\delta
\end{equation}
and
\begin{equation}
\theta^{n}\cdot\theta'\thickapprox\theta\cdot\theta'\thickapprox\frac{1}{n+1}\delta,
\end{equation}
both of which prove to be fully consistent from a mathematical point
of view. Hence, generally speaking, it can be concluded that

\begin{equation}
\theta\cdot\theta'\thickapprox\theta\cdot\delta\thickapprox A\delta
\end{equation}
applies for some constant $A$, which makes it perfectly clear that
it would be both wrong and misleading to naively conclude that in
the Colombeau algebra $\theta(x)$ times $\delta(x)$ is just $\frac{1}{2}\delta(x)$.
Instead, as it turns out, association enables one to model $\theta(x)$
times $\delta(x)$ in a large number of ways, which shows that the
problem of modeling the product $\theta(x)\cdot\delta(x)$ is more
sophisticated than one would expect at first glance. The same applies
to the residual products depicted in $(8)$, where once more the choice
of the mollifier proves to be decisive for modeling the latter. Association
relations $(8)$ are thus to be understood in a generalized sense,
making it clear that is generally not possible the constants $A$
and $B$ in a unique and unambiguous manner. By a suitable choice
of regularization, however, it can be ensured that both constants
$A$ and $B$ are different from zero. See here \cite{grosser2001geometric}
for further details.

Now that this has been clarified, let the following be taken into
consideration: Given an ambient spacetime $(\mathcal{M},g)$ and two
spacetime partitions $(\mathcal{M}^{\pm},g^{\pm})$ with $\mathcal{M}^{+}\cup\mathcal{M}^{-}\subseteq\mathcal{M}$,
the existence of a generalized function $(\vartheta{}_{\epsilon})_{\epsilon}(x)$
with the properties mentioned above allows for the definition of the
generalized ambient metric $g_{ab}(x)$ resulting from the regularized
expression

\begin{equation}
g_{\epsilon ab}=\vartheta_{\epsilon}\,g_{ab}^{+}+(1-\vartheta_{\epsilon})\,g_{ab}^{-}.
\end{equation}
This ambient metric is thus defined in such a way that
\begin{equation}
\underset{\epsilon\rightarrow0}{\lim}\langle g_{\epsilon ab},\nu_{\varphi}\rangle\equiv\langle g_{ab},\nu_{\varphi}\rangle\equiv\langle\theta g_{ab}^{+},\nu_{\varphi}\rangle+\langle(1-\theta)g_{ab}^{-},\nu_{\varphi}\rangle
\end{equation}
and thus

\begin{equation}
\underset{\epsilon\rightarrow0}{\lim}\,g_{\epsilon ab}\approx g_{ab}\approx\theta g_{ab}^{+}+(1-\theta)g_{ab}^{-}
\end{equation}
applies in a generalized sense. 

Given that $\nabla_{a}\vartheta_{\epsilon}^{\pm}=\pm\zeta_{a}\,\vartheta'_{\epsilon}$
applies for some co-vector field $\zeta_{a}$, the corresponding generalized
Christoffel symbols can be determined to be
\begin{equation}
\Gamma_{\,\epsilon bc}^{a}=\vartheta_{\epsilon}\,\Gamma_{\,bc}^{+a}+(1-\vartheta_{\epsilon})\,\Gamma_{\,bc}^{-a}+\vartheta'_{\epsilon}\,\mathcal{C}_{\epsilon bc}^{a},
\end{equation}
provided that $\mathcal{C}_{\epsilon bc}^{a}=\frac{1}{2}([g_{\epsilon b}^{a}]\zeta_{c}+[g_{\epsilon\;c}^{a}]\zeta_{b}-\zeta_{\epsilon}^{a}[g_{bc}])$
with $\zeta_{\epsilon}^{a}=g_{\epsilon}^{ad}\zeta_{d}$, $[g_{\epsilon b}^{a}]=g_{\epsilon}^{ad}[g_{db}]$
and $[g_{bc}]\equiv g_{bc}^{+}-g_{bc}^{-}$ is assumed to apply in
this context. A natural requirement by the thin shell formalism is
that the condition

\begin{equation}
\mathcal{C}_{bc}^{a}\approx\underset{\epsilon\rightarrow0}{\lim}\,\mathcal{C}_{\epsilon bc}^{a}\approx0
\end{equation}
is met, which is tantamount to requiring that

\begin{equation}
\underset{\epsilon\rightarrow0}{\lim}\,\Gamma_{\,\epsilon bc}^{a}\approx\Gamma_{\,bc}^{a}\approx\theta\,\Gamma_{\,bc}^{+a}+(1-\theta)\,\Gamma_{\,bc}^{-a}
\end{equation}
is satisfied. 

With that clarified, it becomes clear that the associated distribution-valued
Riemann tensor $R_{\,bcd}^{a}$ of the ambient geometry $(\mathcal{M},g)$
can be calculated by considering the Colombeau limit of the regularized
expression

\begin{equation}
R_{\,\epsilon bcd}^{a}=\vartheta_{\epsilon}\,R_{\,bcd}^{+a}+(1-\vartheta_{\epsilon})\,R_{\,bcd}^{-a}+\vartheta'_{\epsilon}\,H_{\,bcd}^{a}+\vartheta''_{\epsilon}\,\mathcal{H}_{\,\epsilon bcd}^{a}
\end{equation}
by virtue that the definitions $H_{\,bcd}^{a}=2[\Gamma_{\,b[d}^{a}]\zeta_{c]}$
and $\mathcal{H}_{\,\epsilon bcd}^{a}=2\mathcal{C}_{\,\epsilon b[d}^{a}\zeta_{c]}$
with $[\Gamma_{\,bc}^{a}]\equiv\Gamma_{\,bc}^{+a}-\Gamma_{\,bc}^{-a}$
are used in that regard. Due to the validity of $(8)$, this yields 

\begin{equation}
\underset{\epsilon\rightarrow0}{\lim}\,R_{\,\epsilon bcd}^{a}\approx R_{\,bcd}^{a}\approx\theta\,R_{\,bcd}^{+a}+(1-\theta)\,R_{\,bcd}^{-a}+\delta\,H_{\,bcd}^{a}
\end{equation}
in the limit $\epsilon\rightarrow0$. Using the decomposition 
\begin{equation}
[\Gamma_{\,bc}^{a}]=\gamma_{b}^{\,a}\zeta_{c}+\gamma_{c}^{\,a}\zeta_{b}-\gamma_{bc}\zeta^{a}
\end{equation}
for the difference tensor $[\Gamma_{\,bc}^{a}]$, where $\gamma_{ab}$
is a symmetric tensor field to be further specified below, it is not
hard to check that the tensor field $H_{\,bcd}^{a}$ occurring in
relation $(20)$ can be determined to be

\begin{equation}
H_{\,bcd}^{a}=\frac{1}{2}(\gamma_{\,d}^{a}\zeta_{b}\zeta_{c}-\gamma_{\,c}^{a}\zeta_{b}\zeta_{d}+\gamma{}_{bc}\zeta^{a}\zeta_{d}-\gamma_{bd}\zeta^{a}\zeta_{c}).
\end{equation}
Given this result, Einstein's field equations can ultimately be formulated
in a distributional sense. To do so, it may be taken note of the fact
that also the Einstein tensor splits into three parts, i.e.

\begin{equation}
\underset{\epsilon\rightarrow0}{\lim}\,G_{\,\epsilon b}^{a}\approx G_{\,b}^{a}\approx\theta\,G_{\,b}^{+a}+(1-\theta)\,G_{\,b}^{-a}+\delta\,\rho_{\;b}^{a},
\end{equation}
where $\rho_{\;b}^{a}=H_{\;b}^{a}-\frac{1}{2}\delta_{\;b}^{a}H$ (with
$H\equiv g^{ab}\vert_{\Sigma}H_{ab}$) applies by definition. The
stress-energy tensor has to be of the form, i.e.

\begin{equation}
\underset{\epsilon\rightarrow0}{\lim}\,T_{\,\epsilon b}^{a}\approx T_{\,b}^{a}\approx\theta\,T_{\,b}^{+a}+(1-\theta)\,T_{\,b}^{-a}+\delta\,\tau_{\;b}^{a},
\end{equation}
thereby implying that the local relations
\begin{equation}
G_{\,b}^{\pm a}=8\pi T_{\;b}^{\pm a},\;\rho_{\;b}^{a}=8\pi\tau_{\;b}^{a}
\end{equation}
have to be satisfied in order to ensure that the ambient field equations
\begin{equation}
G_{\,b}^{a}\approx8\pi T_{\;b}^{a}
\end{equation}
are consistently met in a distributional sense. Here, as may be noted,
the local stress-energy tensor $T_{ab}^{\pm}(x)$ merely encodes the
local matter content of the local spacetimes $(\mathcal{M}^{\pm},g^{\pm})$,
not that of the entire ambient spacetime $(\mathcal{M},g)$. The latter
is rather encoded in the full-fledged stress-energy tensor $T_{ab}(x)$
of $(\mathcal{M},g)$, which additionally entails the (typically non-vanishing)
difference tensors $\rho_{ab}(x)$ and $\tau_{ab}(x)$. 

Given this set of relations, it becomes apparent that the pair of
spacetime partitions $(\mathcal{M}^{\pm},g^{\pm})$ can smoothly be
joined if and only if
\begin{equation}
H_{\,bcd}^{a}=\rho_{\;b}^{a}=0.
\end{equation}
 applies locally at $\Sigma$. However, irrespective of the causal
structure of said boundary portion, this is tantamount to requiring
that
\begin{equation}
\gamma_{ab}=0;\;\gamma_{ab}\zeta^{b}=\gamma=0
\end{equation}
is satisfied at $\Sigma$, where the semicolon indicates that two
different conditions are depicted in $(28)$.

In combination with relation $(17)$, as may be noted, the first of
these conditions proves identical to the Darmois-Isreal junction conditions
\cite{darmois1927equations,israel1966singular,poisson2004relativist,sen1924grenzbedingungen}

\begin{equation}
[h_{ab}]=[K_{ab}]=0
\end{equation}
in the event that a non-null surface layer is considered, which are
set up with respect to the pair of induced Riemannian three-metrics
$h_{ab}^{\pm}=g_{ab}^{\pm}+\epsilon n_{a}^{\pm}n_{b}^{\pm}$ and extrinsic
curvatures $K_{ab}^{\pm}=\frac{1}{2}h_{a}^{\pm\:c}h_{b}^{\pm\:d}L_{n^{\pm}}h_{cd}^{\pm}$
with $n_{\pm}^{a}$ constituting the corresponding hypersurface normal
vector fields and $L_{n^{\pm}}$ the Lie derivative along these vector
fields. In the alternative case where a mixed or null boundary layer
is considered, the listed conditions can be rewritten in the form
\cite{barrabes1991thin,mars1993geometry}

\begin{equation}
[\mathcal{H}_{ab}]=0;\;[\mathcal{H}_{ab}]\zeta^{b}=[\mathcal{H}]=0.
\end{equation}
To see this, it may be taken into account that $\zeta^{a}(x)$ can
be chosen to be a hypersurface normal vector field meeting the condition
$\nabla^{[a}\zeta^{b]}=\nabla_{[a}\zeta_{b]}=0$, where the corresponding
co-vector field $\zeta_{a}(x)$ can be completed to a co-tetrad of
the form $\{\zeta_{a},e_{a}^{\rho}\}$ with $\{e_{\:a}^{\rho}\}\in T^{*}(\Sigma)$
and $\rho=1,2,3$. This co-tetrad results naturally from a suitable
identification of a pair of local co-tetrads $\{\zeta_{a}^{\pm},e_{a}^{\rho}\}$
across $\Sigma$ and is associated with a so-called rigging vector
field $\xi^{a}(x)$ by the relation $\zeta_{a}\xi^{a}=-1$. This same
vector field can be completed to a tetrad $\{\xi^{a},E_{\:\rho}^{a}\}\in T(\Sigma)$
resulting from the identification of a pair of local tetrads $\{\xi_{\pm}^{a},E_{\:\rho}^{a}\}$
across $\Sigma$. The corresponding basis and co-basis elements have
the properties that $e_{\:a}^{\rho}E_{\:\sigma}^{a}=\delta_{\:\sigma}^{\rho}$,
$e_{\:a}^{\rho}\zeta^{a}=E_{\:\sigma}^{a}\xi_{a}=0$ and $(g_{ab}\vert_{\Sigma}-g_{ab}^{\pm}\vert_{\Sigma})E_{\:\rho}^{a}E_{\:\sigma}^{a}=0$
applies, thereby allowing to set up a projector onto $\Sigma$ of
the type $o_{\,a}^{c}=\delta{}_{\,a}^{c}+\zeta^{c}\xi_{a}$ having
the properties $o_{\,a}^{c}o_{\,b}^{a}=o_{\,b}^{c}$ and $o_{\,a}^{c}\xi^{a}=o_{\,a}^{c}\zeta_{c}=0$.
This projector allows the construction of a two-form $\mathcal{H}_{ab}=2o_{\,a}^{c}o_{\,b}^{d}\nabla_{(c}\xi_{d)}$,
which gives rise to a difference tensor field $[\mathcal{H}_{ab}]\equiv\mathcal{H}_{ab}^{+}-\mathcal{H}_{ab}^{-}$
that proves to be identical with the expression $\gamma_{ab}(x)$
occurring in relations $(21)$ and $(22)$, i.e. $\gamma_{ab}\equiv[\mathcal{H}_{ab}]$. 

As a result, however, it becomes clear that the junctions conditions
$(29)$ and $(30)$ result directly from $(28)$ both in the null
and non-null cases. The null case occurs naturally for the choice
$\xi_{a}\equiv l_{a}$, $\zeta^{a}\equiv k^{a}$ and $o_{\,a}^{c}\equiv\delta{}_{\,a}^{c}+k^{c}l_{a}$,
where $k^{a}(x)$ and $l^{a}(x)$ are normalized null vector fields
in $(\mathcal{M},g)$. The non-null cases arise naturally for the
choice $\xi_{a}\equiv\epsilon n_{a}$, $\zeta^{a}\equiv n^{a}$ and
$o_{\,a}^{c}\equiv h{}_{\,a}^{c}\equiv\delta{}_{\,a}^{c}+\epsilon n^{c}n_{a}$
with $\epsilon=\pm1$, where $h_{ab}(x)$ is a Riemannian three-metric
induced by $g_{ab}(x)$ through the pull-back $\Phi^{*}(g)_{ab}(x)$.
This metric coincides exactly at $\Sigma$ with the local induced
Riemannian three-metrics $h_{ab}^{\pm}=g_{ab}^{\pm}+\epsilon n_{a}^{\pm}n_{b}^{\pm}$,
so that it becomes clear that the first junction condition in $(29)$
is met. Moreover, since in the non-null cases one has $\gamma_{ab}\equiv[\mathcal{H}_{ab}]\equiv\epsilon[K_{ab}]$,
it becomes clear by relations $(20)$ and $(22)$ that local spacetime
partitions $(\mathcal{M}^{\pm},g^{\pm})$ can smoothly be joined across
a timelike/spacelike hypersurfaces if and only if conditions $(27)$
are met. In the null case, on the other hand, there is also the possibility
that $(\mathcal{M}^{\pm},g^{\pm})$ can be smoothly joined across
a null hypersurfaces if the latter type of conditions in $(30)$ are
met. Thus, it becomes clear that the approach used proves fully consistent
with the mixed (generalized) thin shell formalism discussed in \cite{mars1993geometry,racsko2021variational}.
In many respects, though, as shown en detail only in the next section
of this work, the adopted approach proves to be mathematically more
rigorous and flexible than the conventional methods used in the literature.
This is only explained in detail below, in the second and concluding
section of this paper.

\section{Generalized Energy Conditions, Nonlinear Curvature Terms and Singular
Matching}

In the preceding section, it was shown that the geometric framework
of the mixed thin-shell formalism results naturally from the generalized
formalism in the Colombeau limit $\epsilon\rightarrow0$. Building
on this, the following section will show to what extent the method
used offers practical advantages in the calculation of physical quantities
relevant to general relativity and higher-derivative theories of gravity,
as well as for matching specific types of singular spacetimes with
distributional metrics. 

To begin with, it shall first be demonstrated that the dominant energy
condition can be formulated for thin shells. To this end, let the
generalized vector field $j^{a}\equiv T_{\;b}^{a}v^{b}\approx\underset{\epsilon\rightarrow0}{\lim}\,T_{\epsilon\:b}^{a}v^{b}$
be considered. Given a suitable fixed choice of regularization, the
four-product 

\begin{align}
j_{a}j^{a}\equiv T_{\;b}^{a} & T_{\;a}^{c}v^{b}v_{c}\approx\underset{\epsilon\rightarrow0}{\lim}\,T_{\epsilon\:b}^{a}T_{\epsilon\:a}^{c}v^{b}v_{c}\approx\\
\{\theta\,T_{+\,b}^{a}T_{+\,a}^{c}+ & (1-\theta)\,T_{-\,b}^{a}T_{-\,a}^{c}+\delta\,S_{\;b}^{a}S_{\;a}^{c}\}v^{b}v_{c},\nonumber 
\end{align}
can then be calculated, provided that the definition
\begin{equation}
S_{\;c}^{a}S_{\;b}^{c}\equiv B\tau_{\;c}^{a}\tau_{\;b}^{c}+2AT_{+\,c}^{a}\tau_{\;b}^{c}+2(1-A)T_{-\,c}^{a}\tau_{\;b}^{c}
\end{equation}
is used in the present context. Given this result, which apparently
includes no ill-defined products of the delta distribution, the dominant
energy condition for thin shells can be formulated to be

\begin{equation}
j_{a}j^{a}\leq0,
\end{equation}
where it may be required for the sake of convenience that

\begin{equation}
T_{\pm\,b}^{a}T_{\pm\,a}^{c}v^{b}v_{c}\leq0,\;S_{\;b}^{a}S_{\;a}^{c}v^{b}v_{c}\leq0.
\end{equation}
Thus, the dominant energy condition gives rise to a generalized scalar
field that is perfectly well-defined from a distributional point of
view; although its concrete form is strongly dependent of the choice
of regularization resp. the choice of the Colombeau algebra of manifold-valued
generalized functions used. This freedom of choice is reflected by
the fact that the relevant constants $A$ and $B$ can only be fixed
by a concrete choice of the Colombeau objects $(\vartheta_{\epsilon}(x))_{\epsilon}$,
and $(\vartheta'_{\epsilon}(x))_{\epsilon}$ considered. It is worth
noting here that a comparable result has already been derived in \cite{huber2020junction}
for the special case of generalized Kerr-Schild and Gordon metrics.
However, the general case considered in this paper has, as it seems,
not yet been considered in the literature on the subject so far. 

With this clarified, a similar picture emerges in the calculation
of various curvature expressions, some of which play a role in Einstein-Hilbert
gravity, but above all in higher-derivative theories of gravitation.
Relevant curvature invariants, which can be calculated in a generalized
sense, are

\begin{align}
\underset{\epsilon\rightarrow0}{\lim}\,R_{\epsilon abcd}R_{\epsilon}^{abcd} & \approx R_{abcd}R^{abcd}\approx\theta\,R_{abcd}^{+}R_{+}^{abcd}+(1-\theta)\,R_{abcd}^{-}R_{-}^{abcd}+\\
+ & \delta\,O,\nonumber \\
\underset{\epsilon\rightarrow0}{\lim}\,R_{\epsilon ab}R_{\epsilon}^{ab} & \approx R_{ab}R^{ab}\approx\theta\,R_{ab}^{+}R_{+}^{ab}+(1-\theta)\,R_{ab}^{-}R_{-}^{ab}\\
+ & \delta\,U,\nonumber \\
\underset{\epsilon\rightarrow0}{\lim}\,R_{\epsilon}^{2} & \approx R^{2}\approx\theta\,R^{+2}+(1-\theta)\,R^{-2}+\delta\,V,
\end{align}
giving rise to the smooth scalar fields
\begin{align}
O & =BH_{abcd}H^{abcd}+A(R_{abcd}^{+}H^{abcd}+H_{abcd}R_{+}^{abcd})+\\
 & +(1-A)(R_{abcd}^{-}H^{abcd}+H_{abcd}R_{-}^{abcd}),\nonumber \\
U & =BH_{ab}H^{ab}+A(R_{ab}^{+}H^{ab}+H_{ab}R_{+}^{ab})+\\
 & +(1-A)(R_{ab}^{-}H^{ab}+H_{ab}R_{-}^{ab}),\nonumber \\
V & =BH^{2}+2AR^{+}H+2(1-A)R^{-}H.
\end{align}
The resulting expressions, as may be noted, again do not contain squares
of the delta distribution or other ill-defined terms. In this sense,
the proposed method proves to be advantageous compared to traditional
distributional approaches predominantly used in literature, which
have to deal with such undefined terms. A similar approach to handling
nonlinear curvature terms, as is to be noted, was already adopted
in \cite{foukzon2020singular}, for the purpose of calculating curvature
quantities and their nonlinear invariants for the distributional Schwarzschild
geometry by using methods from Colombeau theory.

That said, to pick out just one concrete example in which the derived
results prove useful, one may take a look at quadratic theories of
gravity described by the Lagrangian density
\begin{equation}
\mathcal{L}=\frac{1}{16\pi}(R-2\Lambda+a_{1}R^{2}+a_{2}R_{ab}R^{ab}+a_{3}R_{abcd}R^{abcd})+\mathcal{L}_{M},
\end{equation}
where $\mathcal{L}_{M}$ is the matter Lagrangian density. In the
standard smooth theory, a variation of this Lagrangian leads to the
higher-order field equations
\begin{equation}
G_{ab}+\Lambda g_{ab}+\mathfrak{G}_{ab}=8\pi T_{ab},
\end{equation}
which extend Einstein-Hilbert gravity based on the system of correction
terms
\begin{align}
\mathfrak{G}_{ab} & =2\{a_{1}RR_{ab}-2a_{3}R_{ac}R_{\,b}^{c}+a_{3}R_{acde}R_{b}^{\:cde}+(a_{2}+2a_{3})R_{acbd}R^{cd}\\
 & -(a_{1}+\frac{1}{2}a_{2}+a_{3})\nabla_{a}\nabla_{b}R+(\frac{1}{2}a_{2}+2a_{3})\oblong R_{ab}\}\nonumber \\
 & -\frac{1}{2}g_{ab}\{a_{1}R^{2}+a_{2}R_{ab}R^{ab}+a_{3}R_{abcd}R^{abcd}-(4a_{1}+a_{2})\oblong R\}.\nonumber 
\end{align}
In the singular case, in which the Riemann curvature tensor, the Ricci
tensor and the Ricci scalar are tensor distributions resulting from
$(20)$, it is pretty much unclear at first sight whether expressions
$(41)$ and $(42)$ constitute well-defined scalar and tensor distributions.
As relations $(35-40)$ show, however, at least the Lagrangian density
$(41)$ as well as parts of $(43)$ are perfectly well-defined from
a distributional point of view. Moreover, using the relation $(19)$
in conjunction with $(2),$ $(8)$ and $(17)$, it becomes clear that
also the expressions

\begin{align}
\underset{\epsilon\rightarrow0}{\lim}\,R_{\epsilon acde}R_{\epsilon b}^{\:cde} & \approx R_{acde}R_{b}^{\:cde}\approx\theta\,R_{acde}^{+}R_{+b}^{\:cde}+(1-\theta)\,R_{acde}^{-}R_{-b}^{\:cde}+\\
+ & \delta\,W_{ab},\nonumber \\
\underset{\epsilon\rightarrow0}{\lim}\,R_{\epsilon acbd}R_{\epsilon}^{cd} & \approx R_{acbd}R^{cd}\approx\theta\,R_{abcd}^{+}R_{+}^{bd}+(1-\theta)\,R_{abcd}^{-}R^{bd}\\
+ & \delta\,X_{ab},\nonumber \\
\underset{\epsilon\rightarrow0}{\lim}\,R_{\epsilon ac}R_{\epsilon\,b}^{c} & \approx R_{ac}R_{\,b}^{c}\approx\theta\,R_{ac}^{+}R_{+\,b}^{c}+(1-\theta)\,R_{ac}^{-}R_{-\,b}^{c}+\delta\,Y_{ab},\\
\underset{\epsilon\rightarrow0}{\lim}\,R_{\epsilon}R_{\epsilon ab} & \approx RR_{ab}\approx\theta\,R^{+}R_{ab}^{+}+(1-\theta)\,R^{-}R_{ab}^{-}+\delta\,Z_{ab},
\end{align}
consitute well-defined tensor distributions, which are defined with
respect to the smooth tensor fields
\begin{align}
W_{ab} & =BH_{acde}H_{b}^{\:cde}+A(R_{acde}^{+}H_{b}^{\:cde}+H_{acde}R_{+b}^{\:cde})+\\
 & +(1-A)(R_{acde}^{-}H_{b}^{\:cde}+H_{acde}R_{-b}^{\:cde}),\nonumber \\
X_{ab} & =BH_{acbd}H^{cd}+A(R_{abcd}^{+}H^{cd}+H_{acbd}R_{+}^{cd})+\\
 & +(1-A)(R_{abcd}^{-}H^{cd}+H_{acbd}R_{-}^{cd}),\nonumber \\
Y_{ab} & =BH_{ac}H_{\,b}^{c}+A(R_{ac}^{+}H_{\,b}^{c}+H_{ac}R_{+\,b}^{c})\\
 & +(1-A)(R_{ac}^{-}H_{\,b}^{c}+H_{ac}R_{-\,b}^{c}),\nonumber \\
Z_{ab} & =BHH_{ab}+A(R_{+}H_{ab}+HR_{ab}^{+})\\
 & +(1-A)(R_{-}H_{ab}+HR_{ab}^{-}).\nonumber 
\end{align}
This makes it clear, however, that also the correction field $(43)$
constitutes a well-defined tensor distribution. Yet, compared to previous
results on the subject \cite{berezin2020double,ivanova2021null,reina2016junction,senovilla2013junction,senovilla2015double,senovilla2018equations},
the decisive difference is that again no undefined delta-squared terms
arise at any point of the calculation and therefore no dubious cancellations
of the same due to suitable junction conditions need to be contemplated.
Still, as may be checked, both the conventional Darmois-Israel and/or
Mars-Senovilla junction conditions as well as the generalized junction
conditions derived in \cite{reina2016junction} result directly from
expressions $(43)$ and related expressions $(44-51)$ calculated
above. The applied method of collecting quadratic terms in the Lagrangian
$(41)$ in a common-factor fashion thus seems to work exactly as expected
(for further details, the reader is here referred to \cite{reina2016junction}).
This shows that Colombeau's theory of generalized functions proves
useful for dealing with thin shells in quadratic gravity, and may
even prove similarly useful in dealing with other high-derivative
theories of gravity.

Having outlined these applications of the generalized approach to
the thin-shell formalism discussed above, the next step will be to
make clear that said approach also lends itself to gluing together
different classes of singular spacetimes with distributional metrics
(containing a delta distribution term). The latter is of interest
because so far only smooth metrics could be matched with the 'conventional'
distributional methods of the thin-shell formalism.

To show that a little bit more is possible, however, let again an
ambient spacetime $(\mathcal{M},g)$ with local partitions $(\mathcal{M}^{\pm},g^{\pm})$
be considered. Assuming that the pair of local metrics $g_{ab}^{\pm}$
and their inverses $g_{\pm}^{ab}$ have Colombeau counterparts $g_{\epsilon ab}^{\pm}$
and $g_{\epsilon\pm}^{ab}$ that can be decomposed into the form 
\begin{equation}
g_{\epsilon ab}^{\pm}=g_{0ab}^{\pm}+\delta_{\epsilon}e_{ab}^{\pm},\;g_{\epsilon\pm}^{ab}=g_{0\pm}^{ab}+\delta{}_{\epsilon}f_{\pm}^{ab},
\end{equation}
it becomes possible to split these fields in a smooth regular parts
and singular parts, i.e. 
\begin{equation}
\underset{\epsilon\rightarrow0}{\lim}\,g_{\epsilon ab}^{\pm}\approx g_{ab}^{\pm}\approx g_{0ab}^{\pm}+\delta\,e_{ab}^{\pm},\;\underset{\epsilon\rightarrow0}{\lim}\,g_{\epsilon\pm}^{ab}\approx g_{\pm}^{ab}\approx g_{0\pm}^{ab}+\delta\,f_{\pm}^{ab},
\end{equation}
where a smooth generalized function meeting the condition $\delta{}_{\epsilon}\equiv\vartheta'_{\epsilon}$
has been considered in order to set up $(43)$. The exact form of
the ambient metric $g_{ab}$ then follows from the system of association
relations
\begin{align}
\underset{\epsilon\rightarrow0}{\lim}\,g_{\epsilon ab} & \approx g_{ab}\approx\theta\,g_{0ab}^{+}+(1-\theta)\,g_{0ab}^{-}+\delta\,e_{ab},\\
\underset{\epsilon\rightarrow0}{\lim}\,g_{\epsilon}^{ab} & \approx g^{ab}\approx\theta\,g_{0+}^{ab}+(1-\theta)\,g_{0-}^{ab}+\delta\,f^{ab},\nonumber 
\end{align}
provided that the definitions $e_{ab}\equiv Ae_{ab}^{+}+(1-A)e_{ab}^{-}$
and $f_{ab}\equiv Af_{ab}^{+}+(1-A)f_{ab}^{-}$ is used in the present
context. To make matters a little simpler right from the outset, let
it be assumed that $g_{0ab}^{+}\equiv g_{0ab}\equiv g_{0ab}^{-}$
is satisfied, so that $g_{ab}\approx g_{0ab}+\delta\,e_{ab}$ and
$g^{ab}\approx g_{0}^{ab}+\delta\,f^{ab}$ applies for the sake simplicity.
Thus, by the requirement that $g_{ac}g^{cb}\approx\delta_{a}^{\;b}$
is satisfied in the distributional sense, it becomes clear that 

\begin{equation}
e_{\;b}^{a}+f_{\;b}^{a}+Be_{\;c}^{a}f_{\;b}^{c}=0
\end{equation}
needs to apply for reasons of consistency. By rewriting $(54)$, it
is found then that the difference of the regularized ambient and background
Levi-Civita connections $\Gamma_{\epsilon\,bc}^{a}$ and $\Gamma_{0\,bc}^{a}$
gives rise to the generalized difference tensor field

\begin{align}
C_{\epsilon\,bc}^{a} & =\Gamma_{\epsilon\,bc}^{a}-\Gamma_{0\,bc}^{a}=\\
 & \frac{1}{2}(g_{0}^{ad}+\delta{}_{\epsilon}f^{ad})[\nabla_{b}^{0}(e_{dc}\delta{}_{\epsilon})+\nabla_{c}^{0}(e_{bd}\delta{}_{\epsilon})-\nabla_{d}^{0}(e_{bc}\delta{}_{\epsilon})],\nonumber 
\end{align}
where $\nabla_{b}^{0}$ is the covariant derivative with respect to
$g_{0ab}$. This can be used to set up the regularized Riemann tensor
\begin{equation}
R_{\epsilon\,bcd}^{a}=R_{0\,bcd}^{a}+\mathcal{C}_{\epsilon\,bcd}^{a},
\end{equation}
where the definition
\begin{equation}
\mathcal{C}_{\epsilon\,bcd}^{a}=\nabla_{c}^{0}C_{\epsilon\,bd}^{a}-\nabla_{d}^{0}C_{\epsilon\,bc}^{a}+C_{\epsilon\,ec}^{a}C_{\epsilon\,bd}^{e}-C_{\epsilon\,ed}^{a}C_{\epsilon\,bc}^{e}
\end{equation}
has been used. This, in turn, can be used to to set up the corresponding
expressions for the Ricci tensor $R_{\epsilon bd}\equiv R_{\epsilon\,bad}^{a}$
and and the Ricci scalar $R_{\epsilon}\equiv g_{\epsilon}^{bd}R_{\epsilon bd}$
of the geometry. As can readily be seen, these expressions only lead
to meaningful tensor distributions in the Colombeau limit if the condition
\begin{equation}
\underset{\epsilon\rightarrow0}{\lim}\,(C_{\epsilon\,ec}^{a}C_{\epsilon\,bd}^{e}-C_{\epsilon\,ed}^{a}C_{\epsilon\,bc}^{e})\approx0
\end{equation}
is fulfilled. Furthermore, the condition
\begin{equation}
\underset{\epsilon\rightarrow0}{\lim}\,\delta{}_{\epsilon}f^{ac}R_{\epsilon bc}\approx\,\delta f^{ac}R_{bc}\approx\,\delta f^{ac}R_{0bc}
\end{equation}
must be met as a minimum requirement in order to avoid the emergence
of unspecified delta-square terms or something similarly ill-defined
in the field equations of the theory. Once these conditions are met,
however, it is straightforward to set up distributional mixed Einstein
equations of the form $(26)$ using the association relations
\begin{equation}
\underset{\epsilon\rightarrow0}{\lim}\,G_{\epsilon\:b}^{a}\approx\underset{\epsilon\rightarrow0}{\lim}\,(R_{\epsilon\:b}^{a}-\delta_{\:b}^{a}R_{\epsilon})\approx G_{\:b}^{a},\;\underset{\epsilon\rightarrow0}{\lim}\,T_{\epsilon\:b}^{a}\approx T_{\:b}^{a}.
\end{equation}
Now, at first glance, one would expect it to be difficult (if not
impossible) to fulfill conditions $(59)$ and $(60)$ and thus to
formulate mixed field equations in which the delta distribution only
occurs in linear form. But, as already shown in \cite{huber2020junction}
and in previous related work on the subject \cite{huber2020distributional},
there is by all means a class of spacetimes for which these conditions
are satisfied: the class of generalized Kerr-Schild spacetimes. In
view of the geometric setting outlined above, these are ambient spacetimes
with distributional metrics of the form

\begin{align}
\underset{\epsilon\rightarrow0}{\lim}\,g_{\epsilon ab} & \approx g_{ab}\approx g_{0ab}+\delta fk_{a}k_{b},\\
\underset{\epsilon\rightarrow0}{\lim}\,g_{\epsilon}^{ab} & \approx g^{ab}\approx g_{0}^{ab}-\delta fk^{a}k^{b},\nonumber 
\end{align}
according to which, of course, $g_{\epsilon ab}=g_{0ab}+\delta{}_{\epsilon}fk_{a}k_{b}$,
$g_{\epsilon}^{ab}=g_{0}^{ab}-\delta{}_{\epsilon}fk^{a}k^{b}$ and
$k_{a}=g_{0ab}k^{b}$ as well as $f(x)=\theta(x)f_{+}(x)+(1-\theta(x))f_{-}(x)$
applies by definition. As may be noted, a further restriction for
spacetimes of this class is that $(k\nabla)k^{a}=\psi k^{a}$ applies,
where $\psi(x)$ is again a smooth scalar function. By this restriction,
it is ensured that both the mixed Ricci and the mixed Einstein tensor
are linear in the so-called profile function $f(x)\delta(x)$ of the
geometry. That is to say, one finds that the mixed Ricci tensor takes
the form
\begin{align}
R_{\;b}^{a} & =R_{0\:b}^{a}-\frac{1}{2}fR_{0\:c}^{a}k^{c}k_{b}-\frac{1}{2}fR_{0\:b}^{c}k_{c}k^{a}+\\
 & +\frac{1}{2}(\nabla_{c}^{0}\nabla_{0}^{a}(\delta fk^{c}k_{b})+\nabla_{0}^{c}\nabla_{b}^{0}(\delta fk_{c}k^{a})-\nabla_{c}^{0}\nabla_{0}^{c}(\delta fk^{a}k_{b})),\nonumber 
\end{align}
and Ricci scalar reads
\begin{equation}
R=R_{0}-fR_{0\:c}^{d}k_{d}k^{c}+\nabla_{d}^{0}\nabla_{0}^{c}(\delta fk^{d}k_{c})
\end{equation}
which ultimately implies that the mixed Einstein tensor reads
\begin{equation}
G_{\;b}^{a}=R_{\;b}^{a}-\frac{1}{2}\delta{}_{\;b}^{a}R=G_{0\:b}^{a}+\rho{}_{\;b}^{a}
\end{equation}
with
\begin{align}
\rho{}_{\;b}^{a} & =-\frac{1}{2}\delta fR_{0\:c}^{a}k^{c}k_{b}-\frac{1}{2}\delta fR_{0\:b}^{c}k_{c}k^{a}+\frac{1}{2}\delta{}_{\;b}^{a}(\delta fR_{0\:c}^{d}k_{d}k^{c})-\\
 & \frac{1}{2}\delta{}_{\;b}^{a}\nabla_{d}^{0}\nabla_{0}^{c}(\delta fk^{d}k_{c}))+\frac{1}{2}(\nabla_{c}^{0}\nabla_{0}^{a}(\delta fk^{c}k_{b})+\nabla_{0}^{c}\nabla_{b}^{0}(\delta fk_{c}k^{a})-\nabla_{c}^{0}\nabla_{0}^{c}(\delta fk^{a}k_{b})).\nonumber 
\end{align}
Using $(8)$, it thus becomes clear that 
\begin{equation}
f\delta\equiv f_{+}\theta\delta+f_{-}(1-\theta)\delta\approx(Af_{+}+(1-A)f_{-})\delta
\end{equation}
is satisfied. However, this makes it clear that (apart from the ambiguity
of making an appropriate choice for $A$ by selecting a suitable regularization),
the problem of joining generalized Kerr-Schild spacetimes with metrics
and inverse metrics the form 
\begin{equation}
g_{ab}^{\pm}\approx g_{0ab}+\delta\,f_{\pm}k_{a}k_{b},\;g_{\pm}^{ab}\approx g_{0}^{ab}+\delta\,f_{\pm}k^{a}k^{b}
\end{equation}
constitutes a perfectly well-defined matching problem. If, in addition,
a generalized Kerr-Schild ansatz of the form $(53)$ is used for the
matching and it is required that the conditions $\nabla_{[a}k_{b]}=0$
and
\begin{equation}
(k\nabla_{0})(\delta f)\approx0,\;\nabla_{0}^{e}(\delta f)k_{[c}\nabla_{d]}^{0}k_{e}\approx k_{[d}(k\nabla_{0})\nabla_{c]}^{0}(\delta f)\approx0
\end{equation}
are fulfilled, the results of \cite{huber2020distributional} show
that not only the mixed Ricci and Einstein tensors prove to be linear
in the profile function $f(x)\delta(x)$ of the geometry, but also
the Riemann tensor as well as the Ricci and Einstein tensors with
lowered and raised indices. The field equations of the ambient spacetime
$(\mathcal{M},g)$ can then, as also follows from \cite{huber2020distributional},
be derived from an action principle in the same way as those of the
two local spacetime partitions $(\mathcal{M}^{\pm},g^{\pm})$ with
metrics of low regularity. However, this makes it clear that singular
matching is possible in principle; i.e. that it is possible to match
$C^{1,1}$-metrics of generalized Kerr-Schild type across a joint
boundary hypersurface $\Sigma$.

Now that this has been clarified, it may ultimately be noted that
the geometric framework of the thin-shell formalism can be readily
generalized in many respects. This becomes immediately clear once
the following fact is taken into account: A simple way to generalize
the model is to simply replace the generalized function $(\vartheta_{\epsilon})_{\epsilon}(x)$
occurring in ansatz $(13)$ by another generalized function $(\chi_{\epsilon})_{\epsilon}(x)$
that is associated to another Schwartz distribution. This leads to
a completely different matching approach characterized by an ambient
metric of the form

\begin{equation}
\underset{\epsilon\rightarrow0}{\lim}\,g_{\epsilon ab}\approx g_{ab}\approx\chi g_{ab}^{+}+(1-\chi)g_{ab}^{-},
\end{equation}
defined with respect to some generalized function $\chi(x)$ that
does not coincide with a Heaviside step function $\theta(x)$ in the
Colombeau limit $\epsilon\rightarrow0$ but rather with another Schwartz
distribution. A good example for such a generalized function would
be, for instance, one that gives a ramp (generalized) function $x\theta(x)$
in the limit $\epsilon\rightarrow0$, i.e. $\underset{\epsilon\rightarrow0}{\lim}\,\chi_{\epsilon}(x)\approx\chi(x)\equiv x\theta(x)$.
This is because the adopted approach for the given choice leads to
a thick rather than a thin matter shell, which is no longer defined
by a confined stress-energy tensor with compact support on a single
hypersurface of spacetime. Rather, parts of the support of the stress-energy
tensor of the ambient spacetime lie here in an extended transition
region constituting a four-dimensional Lorentzian submanifold of spacetime.
In this way, however, a completely different matching problem is envisaged,
by means of which certain problems of the spacetime thin shell gluing
approach can possibly be overcome. To give one further example of
a generalized function leading to a different matching approach, let
the case be considered in which $\chi_{\epsilon}(x)$ gives a boxcar
(characteristic) function in the limit $\epsilon\rightarrow0$, i.e.
$\underset{\epsilon\rightarrow0}{\lim}\,\Upsilon_{\epsilon}(x;a,b)\equiv\underset{\epsilon\rightarrow0}{\lim}\,\vartheta_{\epsilon}(x-a)-\underset{\epsilon\rightarrow0}{\lim}\,\vartheta_{\epsilon}(x-b)\approx\theta(x-a)-\theta(x-b)\equiv\Upsilon_{a,b}(x)\equiv\chi(x)$.
The latter gives, in the special case $a=b=\frac{1}{2}$, the rectangular
function $\chi(x)\equiv\Upsilon_{-\frac{1}{2},\frac{1}{2}}(x)\equiv\Upsilon(x;-\frac{1}{2},\frac{1}{2})\equiv\theta(x+\frac{1}{2})-\theta(x-\frac{1}{2}).$
This choice then results again in a model that proves more general
than the thin-shell model considered in the present section. The latter
could be used in particular for matching metric pairs $g_{ab}^{\pm}$
of the form $g_{ab}^{+}=g_{0ab}$ and $g_{ab}^{+}=g_{0ab}+e_{ab}$
and thus for the construction of local spacetimes introduced in \cite{huber2020junction}.
Yet, many other useful extensions of the conventional thin-shell formalism
based on Colombeau's theory of algebras of generalized functions are
conceivable, which allow the consideration of more general types of
matching problems than those currently studied in the literature.
As an example, hyperfunctions, a class of generalized functions introduced
by Sato \cite{sato1959theory}, could be used for such a venture,
since the latter are based on the complex calculus and limits of holomorphic
functions and not on duals of spaces of smooth functions. These objects
are of interest since the theory of hyperfunctions includes the theory
Schwartz distributions as a special case and explains, among other
things, special properties of the Dirac delta distribution using Cauchy's
residual theorem. Also, the application of other mathematical techniques
for gluing spacetimes in the general theory of relativity are conceivable,
whereby more powerful models of spacetimes with multiple geometric
structures could be modeled. It seems worth exploring at least some
of these alternative methods in the future.

\section{Physical Examples: Glued pp-Wave and Gravitational Shock Wave Spacetimes}

In the remainder of the paper, the physical significance of the results
derived in the previous section shall be underpinned by concrete examples.
To this end, the gluing of pp-wave and gravitational shock wave spacetimes
is studied in greater detail, thus augmenting previous results on
the subject discussed in \cite{huber2020distributional,huber2020junction}.
Though the results obtained in the previous section are completely
general, the focus in the following will exclusively be placed on
generalized Kerr-Schild spacetimes, as the latter prove to be the
easiest to handle. Moreover, for the sake of simplicity, it will be
assumed that no signature change occurs and that only Lorentzian manifolds
are glued together across their joint boundary.

Concrete examples of singular generalized Kerr-Schild class spacetime
geometries that can be glued together across a joint boundary are
impulsive pp-wave spacetimes such as the Aichelburg-Sexl geometry,
the Lousto-Sanchez geometries and others \cite{aichelburg1971gravitational,barrabes2003lightlike,barrabes2004scattering,hogan2004singular,hotta1993shock,lousto1989gravitational,lousto1990curved,lousto1991gravitational,lousto1992ultrarelativistic,stephani2009exact}.
Further examples are spherically symmetric black hole shock wave geometries
such as the geometries of Dray and 't Hooft and Sfetsos \cite{dray1985gravitational,sfetsos1995gravitational}
as well as their axially symmetric generalizations \cite{huber2021gravitational}.
Additional examples are ultrarelativistic shock wave spacetimes in
cosmological backgrounds such as those constructed in \cite{podolsky2002exact,podolsky2019cut,podolsky2022penrose,podolsky2017penrose,Samann:2023bko}.

To see the latter, let a family of Kerr-Schild spacetimes with a distributional
metric of the form $(62)$ be considered, which is defined with respect
to the flat Minkowskian background metric, so that $g_{0ab}\equiv\eta_{ab}$
applies. Moreover, for the sake of simplicity, let it be assumed that
the profile function of the geometry (which is actually a distribution)
takes the form

\begin{equation}
f(v,y,z)=\delta(v)F(y,z),
\end{equation}
so that the resulting Kerr-Schild ambient spacetime is flat everywhere
except for the null hyperplane $v=0$, where the delta-like impulse
is located. 

A specific representative of this spacetime family is the Aichelburg-Sexl
ultraboost geometry \cite{aichelburg1971gravitational}. The reduced
profile function of this model is given by the expression
\begin{equation}
F(y,z)=c\cdot\ln\sqrt{y^{2}+z^{2}}
\end{equation}
where $c$ is constant. The corresponding spacetime geometry, which
arises (among other things) in the ultrarelativistic limit of a Lorentz-boosted
Schwarzschild geometry in isotropic coordinates, characterizes the
field of a point-like particle close to the speed of light. It constitutes
an exact solution to Einstein's equations, the latter of which reduce
to a single differential equation for the multiplicative part $(72)$
of the profile function, i.e.
\begin{equation}
\Delta_{\mathbb{R}_{2}}F=-16\pi b\cdot\delta^{(2)},
\end{equation}
where $\delta^{(2)}(y,z)=\delta(y)\delta(z)$ applies by definition
and $b$ is constant. In fact, the values calculated in \cite{aichelburg1971gravitational}
for the boosted Schwarzschild geometry are $b=\frac{c}{8}=p$, where
$p$ constitutes the momentum of the considered ultrarelativistic
particle. 

To glue two such Aichelburg-Sexl ultraboost geometries together, as
follows directly from the results of the previous section, a profile
function of the form
\begin{equation}
f=\theta f_{+}+(1-\theta)f_{-}
\end{equation}
can be used; provided that, of course, $f_{\pm}(v,y,z)=c_{\pm}\delta(v)\ln\sqrt{y^{2}+z^{2}}$
applies in this context. Taking into account relation $(67)$ and
additionally choosing $c_{+}=\frac{8p_{1}}{A}$ and $c_{-}=\frac{8p_{2}}{1-A}$
in order to eliminate the dependence of the choice of regularization,
another Aichelburg-Sexl spacetime with profile function $(71)$ results
from the undertaken matching with the values $\frac{c}{8}=p_{1}+p_{2}=:p$
occurring in $(72)$ and $(73)$. The matching of two Aichelburg-Sexl
spacetimes with momenta $p_{1}$ and $p_{2}$ thus yields again an
Aichelburg-Sexl ultraboost geometry with total momentum $p_{1}+p_{2}$;
independet of the choice of regularization used for setting up $(67)$. 

In completely analogous fashion, two Lousto-Sanchez spacetimes \cite{lousto1990curved}
can be glued together; types of spacetimes that belong to a class
of Kerr-Schild geometries with profile distribution $(71)$ with the
corresponding reduced part being given by the expression 
\begin{equation}
F(y,z)=c\ln\sqrt{y^{2}+z^{2}}+\frac{d}{\sqrt{y^{2}+z^{2}}}.
\end{equation}
The given type of pp-wave spacetime arises naturally in the ultrarelativistic
limit of the boosted Reissner-Nordström geometry under the assumption
that the charge $e=\gamma^{\frac{1}{2}}p_{e}$ required for the calculation
vanishes in said limit. It constitutes an exact solution of Einstein's
equations that is flat everywhere except on the null plane, where
the concrete values are with $c=8p$ and $d=\frac{3\pi}{2}p_{e}^{2}$.
Proceeding essentially in same way as above, two partitions of this
type of spacetime can be glued together, which yields a Lousto-Sanchez
spacetime with reduced profile function of the form $(75)$ with $c=p=p_{1}+p_{2}$
and $d=\frac{3\pi}{2}(p_{1e}^{2}+p_{2e}^{2})=:\frac{3\pi}{2}p_{e}^{2}$,
where $p_{j}$ and $p_{je}$ with $j=1,2$ are the momenta of the
two Lousto-Sanchez partitions of spacetime.

Further classes of distributional spacetimes that can be glued together
with methods presented in the previous section are gravitational shock
wave spacetimes such as Dray-'t Hooft geometry and the model of Sfetsos.
The latter are singular generalized Kerr-Schild spacetimes of the
form $(62)$ with a profile distribution of the form

\begin{equation}
f(v,r,\vartheta,\phi)=F(\vartheta,\phi)\delta(v)\delta(r-r_{+}),
\end{equation}
which is given with respect to a reduced profile function $F(\theta,\phi)$
that is a solution of the distributional Einstein equations $(26)$.
The corresponding classes of spacetime geometries, which can be obtained
by a Kerr-Schild deformation of the Schwarzschild and Reissner-Nordström
black hole metrics, characterize the field of a gravitational shock
wave generated by a point-like particle located at the event horizon
of the respective types of black holes. In the more general case of
Sfetsos' spacetime \cite{sfetsos1995gravitational}, the profile $(76)$
can be obtained by solving the differential relation
\begin{equation}
\langle(\Delta_{\mathbb{S}_{2}}F-cF),\varphi\rangle=\langle2\pi d\delta(\cos\vartheta-1),\varphi\rangle,
\end{equation}
where $d=const.$ and $c=2\kappa r_{+}=\frac{2(r_{+}-M)}{r_{+}}$
with $r_{+}=M+\sqrt{M^{2}-e^{2}}$ and $\kappa=\frac{r_{+}-M}{r_{+}^{2}+a^{2}}$
applies by definition. Since $c\underset{r_{+}\rightarrow2M}{\longrightarrow}1$
is fulfilled in the limit $e\rightarrow0$, it is clear that the Dray-'t
Hooft model \cite{dray1985gravitational} is covered as a special
case by the above. As shown in \cite{huber2021gravitational}, relation
$(77)$ corresponds exactly to the Gaussian hypergeometric differential
equation; the most general Fuchsian differential equation with three
regular singular points. Solutions of $(77)$ are therefore Gaussian
hypergeometric functions; whereby the total solution already corresponds
to a gluing of the homogeneous and particular solutions. More specifically,
the solution of $(77)$ takes the form

\begin{equation}
F=\theta F_{1}+(1-\theta)F_{2},
\end{equation}
where $F_{1}(\vartheta,\phi)$ is the homogeneous and $F_{2}(\vartheta,\phi)$
is the particular solution of the Gaussian hypergeometric differential
equation. Accordingly, all one has to do to glue two Sfetsos spacetimes
together is to make an ansatz of the form $(74)$ for the profile
function of the geometry, where $f_{\pm}(v,r,\vartheta,\phi)=c_{\pm}F_{\pm}(\vartheta,\phi)\delta(v)\delta(r-r_{+})$
are the profile functions of the local partitions of the ambient spacetime.
Here, too, the dependency of the regularization resulting from the
modeling of the product $\theta\cdot\delta$ can be removed by rescaling,
that is, a suitable choice for the constants $c_{\pm}$. This little
trick works not least because solutions of $(77)$ are only unique
up to multiplication by a constant. It thus follows that the matching
of two Sfetsos spacetimes is again a Sfetsos spacetime. 

Of course, one could also have glued a Dray-'t Hooft spacetime to
a Sfetsos spacetime. The method would work analogously in said case.
Perhaps a little more interesting, however, is the gluing of an even
more general class of spacetime geometries, which were constructed
in \cite{huber2021gravitational}. The corresponding classes of spacetimes,
which can be obtained by a Kerr-Schild deformation of the Kerr resp.
Kerr-Newman black hole metrics, characterize the fields of gravitational
shock waves generated by a point-like particle located at the event
horizon of the respective types of black holes. The latter thus constitute
exact solutions to Einstein's equations with distributional profile
taking the form

\begin{equation}
f(v,r,\vartheta,\phi)=e^{\kappa v}\Sigma_{+}^{-1}(\vartheta)F(\vartheta,\phi-\omega_{+}v)\delta(r-r_{+})\delta(v),
\end{equation}
provided that the definitions $\Sigma_{+}(\vartheta)=r_{+}+a^{2}\cos^{2}\vartheta$
and $\omega_{+}=\frac{a}{r_{+}^{2}+a^{2}}$ with $r_{+}=M+\sqrt{M^{2}-a^{2}-e^{2}}$
are used in this context. The reduced part $F(\vartheta,\phi-\omega_{+}v)$
of this profile distribution is a solution of the generalized Dray-'t
Hooft relation
\begin{equation}
\langle\boxtimes F,\varphi\rangle=\langle2\pi d\delta(\cos\vartheta-1),\varphi\rangle,
\end{equation}
 where $d=const.$ and $\boxtimes=\Sigma_{+}^{-1}(\Delta_{\mathbb{S}_{2}}+V)$
with $V=\frac{2r_{+}(M-r_{+})}{\Sigma_{+}}$ is a differential operator
resulting from the distributional Einstein equations $(26)$. As in
the case of the Sfetsos model, relation $(80)$ corresponds to a Fuchsian
differential equation, but now with five regular singular points.
Solutions of $(80)$ have been determined in \cite{huber2021gravitational}
using Mallik's companion matrix approach \cite{mallik1998solutions}.
Yet, as shown in \cite{ishkhanyan2019generalized}, the solutions
of any Fuchsian differential equation with five regular singular points
can actually be identified as generalized hypergeometric functions.
A linear combination of the form $(78)$ of such functions $F_{1}(\vartheta,\phi-\omega_{+}v)$
and $F_{2}(\vartheta,\phi-\omega_{+}v)$ thus constitutes a solution
of $(80)$. Consequently, in order to glue two generalized Kerr-Schild
spacetimes with profiles this type together, one may make again an
ansatz of the form $(74)$, where $f_{\pm}(v,r,\vartheta,\phi)=c_{\pm}e^{\kappa v}\Sigma_{+}^{-1}(\vartheta)F_{\pm}(\vartheta,\phi-\omega_{+}v)\delta(r-r_{+})\delta(v)$
are the profile functions of the different local spacetime partitions.
And again, the dependence on the choice of regularization, which results
from the modeling of the product $\theta\cdot\delta$, can be eliminated
by a suitable choice of $c_{\pm}$. Sure enough, Sfetsos' model results
in the limit $a\rightarrow0$ and the Dray-'t Hooft's model arises
as a special case if the combined limits $a,e\rightarrow0$ are taken.
The results of the gluing techniques used for matching the axially
symmetrical models therefore prove to be completely consistent with
the gluing techniques used in the spherically symmetrical case.

As can readily be seen, the method also works analogously in the case
of pp-waves and gravitational shock waves in cosmological backgrounds.
These cases will therefore not be discussed in detail at this point.

Instead, let attention be drawn to the following interesting observation:
All the spacetime geometries considered in this section have in common
that they were derived on the basis of Penrose's well-known cut-and-paste
method \cite{penrose1972geometry}. Given that the above considerations
show, however, that the extended thin-shell formalism also allows
the derivation of said solutions of Einstein's field equations, it
becomes clear that both methods are consistent in the sense that they
yield exactly the same results. Indeed, this is no coincidence: the
gluing of generalized Kerr-Schild metrics with low regularity of the
given type must always lead to the same results as the cut-and-paste
method in the given generalized Kerr-Schild case. This, it should
be noted, is a direct consequence of the validity of relation $(67)$. 

That the theory of lightlike thin shells also provides the same results
as the cut-and-paste method as regards Penrose's original work has
already been recognized in \cite{manzano2021null}. The fact that
methods other than those used in the present work were adopted in
\cite{manzano2021null} to show this indicates that the thin-shell
formalism and the cut-and-paste methods are likely to be closely related.
It is therefore to be expected that both types of matching techniques
will also prove to be fully compatible with each other in other respects
as well. After all, it may even be expected that both methods result
naturally from a more general geometric formalism, the formalism of
null hypersurface data; see here e.g. \cite{manzano2024abstract,mars1993geometry,mars2013constraint}
for further information. The latter, however, clearly needs to be
investigated in greater detail.

\section*{Summary and Conclusion}

Based on Colombeau's theory of generalized functions, an extension
of mixed thin-shell formalism was presented in this work. More specifically,
by using smooth generalized functions rahter than working with Schwartz
distributions directly, it was shown that handling ill-defined products
of distributions can be avoided in the thin shell formalism. Based
on this observation, it was demonstrated that the generalized formalism
discussed allows to set up the dominant energy condition for thin
shells as well as the definition of nonlinear distribution-valued
curvature invariants needed in higher derivative theories of gravity.
Furthermore, it was shown that the formalism allows the gluing of
spacetimes with special metrics of low regularity (generalized Kerr-Schild
metrics containing a delta distribution term) across a joint boundary
hypersurface in spacetime. Ultimately, as a further application, a
connection to Penrose's cut-and-paste method was established by proving
that the results of the method used in this work lead to the same
results as the cut-and-paste approach in selected examples. Possible
extensions of the conventional thin-shell formalism were also discussed,
allowing thick shells to be considered, and explicit examples of generalized
Kerr-Schild spacetimes that can be matched across a joint null boundary
were given.

The results obtained thus once more impressively demonstrate the significance
of Colombeau's theory of generalized functions for the handling of
nonlinear curvature expressions and ensuing problems in general relativity.

\bibliographystyle{plain}
\addcontentsline{toc}{section}{\refname}\bibliography{0C__Arbeiten_litcol2}

\begin{thebibliography}{10}

\bibitem{aichelburg1971gravitational}
{Peter}~{C} Aichelburg and {Roman}~{U} {Sexl}.
\newblock On the gravitational field of a massless particle.
\newblock {\em General {Relativity} and {Gravitation}}, 2(4):303--312, 1971.

\bibitem{balasin2000generalized}
{Herbert} Balasin.
\newblock Generalized {Kerr}-{Schild} metrics and the gravitational field of a
  massless particle on the horizon.
\newblock {\em Classical and {Quantum} {Gravity}}, 17(9):1913, 2000.

\bibitem{barrabes2001detection}
C~Barrabes and PA~Hogan.
\newblock Detection of impulsive light-like signals in general relativity.
\newblock {\em International Journal of Modern Physics D}, 10(05):711--721,
  2001.

\bibitem{barrabes2002impulsive}
C~Barrabes and PA~Hogan.
\newblock Impulsive light-like signals.
\newblock {\em International Journal of Modern Physics A}, 17(20):2746--2746,
  2002.

\bibitem{barrabes2003lightlike}
{C} Barrabes and {P}{A} {Hogan}.
\newblock Lightlike boost of the {Kerr} gravitational field.
\newblock {\em Physical {Review} {D}}, 67(8):084028, 2003.

\bibitem{barrabes2004scattering}
C~Barrab{\`e}s and PA~Hogan.
\newblock Scattering of high-speed particles in the kerr gravitational field.
\newblock {\em Physical {Review} {D}}, 70(10):107502, 2004.

\bibitem{barrabes1991thin}
C~Barrabes and W~Israel.
\newblock Thin shells in general relativity and cosmology: The lightlike limit.
\newblock {\em Physical Review D}, 43(4):1129, 1991.

\bibitem{berezin2020double}
VA~Berezin, VI~Dokuchaev, Yu~N Eroshenko, and AL~Smirnov.
\newblock Double layer from least action principle.
\newblock {\em Classical and Quantum Gravity}, 38(4):045014, 2020.

\bibitem{colombeau2000new}
Jean~Fran{\c{c}}ois Colombeau.
\newblock {\em New generalized functions and multiplication of distributions},
  volume~84.
\newblock North {Holland} {Publishing} {Co.}, {Amsterdam}, 1984.

\bibitem{colombeau2011elementary}
Jean~Fran{\c{c}}ois Colombeau.
\newblock {\em Elementary introduction to new generalized functions}, volume
  113.
\newblock North {Holland} {Publishing} {Co.}, {Amsterdam}, 1985.

\bibitem{darmois1927equations}
Georges Darmois.
\newblock {\em Les {\'e}quations de la gravitation einsteinienne}.
\newblock Number~25. Gauthier-Villars et. cie., 1927.

\bibitem{dray1985gravitational}
{Tevian} Dray and {Gerard} 't~{Hooft}.
\newblock The gravitational shock wave of a massless particle.
\newblock {\em Nuclear {Physics} {B}}, 253:173--188, 1985.

\bibitem{foukzon2020singular}
Jaykov Foukzon, Elena~R Men'Kova, and Alexander~A Potapov.
\newblock Singular general relativity using the colombeau approach. part i.
  distributional schwarzschild geometry from nonsmooth regularization via
  horizon.
\newblock {\em Physics Essays}, 33(2):180--199, 2020.

\bibitem{groah2007shock}
Jeffrey Groah, Blake Temple, and Joel Smoller.
\newblock {\em Shock wave interactions in general relativity: a locally
  inertial Glimm scheme for spherically symmetric spacetimes}.
\newblock Springer, 2007.

\bibitem{grosser2001geometric}
M.~Grosser, M.~Kunzinger, M.~Oberguggenberger, and R.~Steinbauer.
\newblock {\em Geometric theory of generalized functions with applications to
  general relativity}, volume 537.
\newblock Kluwer {Academic} {Publishers}, {Dordrecht}, 2001.

\bibitem{grosser2012global}
{M} {Grosser}, {M} {Kunzinger}, {R} {Steinbauer}, and {J}{A} {Vickers}.
\newblock A {Global} {Theory} of {Algebras} of {Generalized} {Functions} ii:
  tensor distributions.
\newblock {\em New York Journal of Mathematics}, 18:139--199, 2012.

\bibitem{grosser2001foundations}
{Michael} Grosser, {Eva} {Farkas}, {Michael} {Kunzinger}, and {Roland}
  {Steinbauer}.
\newblock {\em On the foundations of nonlinear generalized functions I and II}.

\bibitem{grosser2002global}
{Michael} Grosser, {Michael} {Kunzinger}, {Roland} {Steinbauer}, and
  {James}~{A} {Vickers}.
\newblock A global theory of algebras of generalized functions.
\newblock {\em Advances in {Mathematics}}, 166(1):50--72, 2002.

\bibitem{hogan2004singular}
Peter~A Hogan and Claude Barrabes.
\newblock {\em Singular Null Hypersurfaces in General Relativity: Light-Like
  Signals from Violent Astrophysical Events}.
\newblock World Scientific, 2004.

\bibitem{hotta1993shock}
{M} Hotta and {M} {Tanaka}.
\newblock Shock-wave geometry with nonvanishing cosmological constant.
\newblock {\em Classical and {Quantum} {Gravity}}, 10(2):307, 1993.

\bibitem{huber2020distributional}
Albert Huber.
\newblock Distributional metrics and the action principle of einstein--hilbert
  gravity.
\newblock {\em Classical and Quantum Gravity}, 37(8):085008, 2020.

\bibitem{huber2020junction}
Albert Huber.
\newblock Junction conditions and local spacetimes in general relativity.
\newblock {\em The European Physical Journal C}, 80:1--19, 2020.

\bibitem{huber2021gravitational}
Albert Huber.
\newblock The gravitational field of a massless particle on the horizon of a
  stationary black hole.
\newblock {\em Classical and Quantum Gravity}, 38(8):085006, 2021.

\bibitem{ishkhanyan2019generalized}
Artur Ishkhanyan and Clemente Cesarano.
\newblock Generalized-hypergeometric solutions of the general fuchsian linear
  ode having five regular singularities.
\newblock {\em Axioms}, 8(3):102, 2019.

\bibitem{israel1966singular}
Werner Israel.
\newblock Singular hypersurfaces and thin shells in general relativity.
\newblock {\em Il Nuovo Cimento B (1965-1970)}, 44(1):1--14, 1966.

\bibitem{ivanova2021null}
ID~Ivanova.
\newblock Null shells and double layers in quadratic gravity.
\newblock In {\em Journal of Physics: Conference Series}, volume 2081, page
  012020. IOP Publishing, 2021.

\bibitem{kunzinger1999rigorous}
{Michael} Kunzinger and {Roland} {Steinbauer}.
\newblock A rigorous solution concept for geodesic and geodesic deviation
  equations in impulsive gravitational waves.
\newblock {\em Journal of {Mathematical} {Physics}}, 40(3):1479--1489, 1999.

\bibitem{kunzinger2002foundations}
{Michael} {Kunzinger} and {Roland} {Steinbauer}.
\newblock Foundations of a nonlinear distributional geometry.
\newblock {\em Acta {Applicandae} {Mathematica}}, 71(2):179--206, 2002.

\bibitem{kunzinger2002generalized}
{Michael} Kunzinger and {Roland} {Steinbauer}.
\newblock Generalized pseudo-{Riemannian} geometry.
\newblock {\em Transactions of the {American} {Mathematical} {Society}},
  354(10):4179--4199, 2002.

\bibitem{kunzinger2009sheaves}
Michael Kunzinger, Roland Steinbauer, and James Vickers.
\newblock Sheaves of nonlinear generalized functions and manifold-valued
  distributions.
\newblock {\em Transactions of the American Mathematical Society},
  361(10):5177--5192, 2009.

\bibitem{kunzinger2003intrinsic}
Michael Kunzinger, Roland Steinbauer, and James~A Vickers.
\newblock Intrinsic characterization of manifold-valued generalized functions.
\newblock {\em Proceedings of the London Mathematical Society}, 87(2):451--470,
  2003.

\bibitem{lousto1989gravitational}
{C}{O} Lousto and {N} {S{\'a}nchez}.
\newblock Gravitational shock waves of ultra-high energetic particles on curved
  spacetimes.
\newblock {\em Physics {Letters} {B}}, 220(1-2):55--60, 1989.

\bibitem{lousto1990curved}
CO~Lousto and N~S{\'a}nchez.
\newblock The curved shock wave space-time of ultrarelativistic charged
  particles and their scattering.
\newblock {\em International Journal of Modern Physics A}, 5(05):915--938,
  1990.

\bibitem{lousto1991gravitational}
{C}{O} Lousto and {N} {S{\'a}nchez}.
\newblock Gravitational shock waves generated by extended sources:
  {Ultrarelativistic} cosmic strings, monopoles and domain walls.
\newblock {\em Nuclear {Physics} {B}}, 355(1):231--249, 1991.

\bibitem{lousto1992ultrarelativistic}
{C}{O} Lousto and {N} {S{\'a}nchez}.
\newblock The ultrarelativistic limit of the boosted {Kerr}-{Newman} geometry
  and the scattering of spin-1/2 particles.
\newblock {\em Nuclear {Physics} {B}}, 383(1-2):377--394, 1992.

\bibitem{mallik1998solutions}
Ranjan~K Mallik.
\newblock Solutions of linear difference equations with variable coefficients.
\newblock {\em Journal of mathematical analysis and applications},
  222(1):79--91, 1998.

\bibitem{mansouri1996equivalence}
R~Mansouri and M~Khorrami.
\newblock The equivalence of darmois-israel and distributional method for thin
  shells in general relativity.
\newblock {\em Journal of Mathematical Physics}, 37(11):5672--5683, 1996.

\bibitem{mansouri2000new}
Reza Mansouri and Kourosh Nozari.
\newblock A new distributional approach to signature change.
\newblock {\em General Relativity and Gravitation}, 32:253--269, 2000.

\bibitem{manzano2021null}
Miguel Manzano and Marc Mars.
\newblock Null shells: general matching across null boundaries and connection
  with cut-and-paste formalism.
\newblock {\em Classical and Quantum Gravity}, 38(15):155008, 2021.

\bibitem{manzano2024abstract}
Miguel Manzano and Marc Mars.
\newblock Abstract formulation of the spacetime matching problem and null thin
  shells.
\newblock {\em Physical Review D}, 109(4):044050, 2024.

\bibitem{mars2013constraint}
Marc Mars.
\newblock Constraint equations for general hypersurfaces and applications to
  shells.
\newblock {\em General Relativity and Gravitation}, 45:2175--2221, 2013.

\bibitem{mars1993geometry}
{Marc} Mars and {Jose}~{M}{M} {Senovilla}.
\newblock Geometry of general hypersurfaces in spacetime: junction conditions.
\newblock {\em Classical and {Quantum} {Gravity}}, 10(9):1865, 1993.

\bibitem{penrose1972geometry}
{Roger} Penrose.
\newblock The geometry of impulsive gravitational waves.
\newblock {\em General {Relativity}, {Papers} in {Honour} of {J}{L} {Synge}},
  pages 101--115, 1972.

\bibitem{podolsky2002exact}
Ji{\v{r}}{\'\i} Podolsk{\`y}.
\newblock Exact impulsive gravitational waves in spacetimes of constant
  curvature.
\newblock In {\em Gravitation: Following the Prague Inspiration: A Volume in
  Celebration of the 60th Birthday of Jir{\'\i} Bi{\v{c}}{\'a}k}, pages
  205--246. World Scientific, 2002.

\bibitem{podolsky2019cut}
Ji{\v{r}}{\'\i} Podolsk{\`y}, Clemens S{\"a}mann, Roland Steinbauer, and Robert
  {\v{S}}varc.
\newblock Cut-and-paste for impulsive gravitational waves with $\lambda$: The
  geometric picture.
\newblock {\em Physical Review D}, 100(2):024040, 2019.

\bibitem{podolsky2022penrose}
Ji{\v{r}}{\'\i} Podolsk{\`y} and Roland Steinbauer.
\newblock Penrose junction conditions with $\lambda$: geometric insights into
  low-regularity metrics for impulsive gravitational waves.
\newblock {\em General Relativity and Gravitation}, 54(9):96, 2022.

\bibitem{podolsky2017penrose}
Ji{\v{r}}{\'\i} Podolsk{\`y}, Robert {\v{S}}varc, Roland Steinbauer, and
  Clemens S{\"a}mann.
\newblock Penrose junction conditions extended: Impulsive waves with gyratons.
\newblock {\em Physical Review D}, 96(6):064043, 2017.

\bibitem{podolsky1998impulsive}
{Ji{\v{r}}{\'\i}} Podolsk{\`y} and {Jerry}~{B} {Griffiths}.
\newblock Impulsive waves in de {Sitter} and anti-de {Sitter} spacetimes
  generated by null particles with an arbitrary multipole structure.
\newblock {\em Classical and {Quantum} {Gravity}}, 15(2):453, 1998.

\bibitem{poisson2004relativist}
Eric Poisson.
\newblock {\em A relativist's toolkit: the mathematics of black-hole
  mechanics}.
\newblock Cambridge university press, 2004.

\bibitem{racsko2021variational}
Bence Racsk{\'o}.
\newblock Variational formalism for generic shells in general relativity.
\newblock {\em Classical and Quantum Gravity}, 39(1):015004, 2021.

\bibitem{reina2016junction}
{Borja} Reina, {Jos{\'e}}~{M}{M} {Senovilla}, and {Ra{\"u}l} {Vera}.
\newblock Junction conditions in quadratic gravity: thin shells and double
  layers.
\newblock {\em Classical and {Quantum} {Gravity}}, 33(10):105008, 2016.

\bibitem{Samann:2023bko}
Clemens S\"amann, Benedict Schinnerl, Roland Steinbauer, and Robert \v{S}varc.
\newblock {Cut-and-paste for impulsive gravitational waves with $\Lambda $: the
  mathematical analysis}.
\newblock {\em Lett. Math. Phys.}, 114(2):58, 2024.

\bibitem{sato1959theory}
Mikio Sato.
\newblock Theory of hyperfunctions.
\newblock {\em J. Fac. Sci. Univ. Tokyo Sec. I}, 8:139--193, 1959.

\bibitem{schwartz1954limpossibilite}
Laurent Schwartz.
\newblock Sur l'impossibilite de la multiplication des distributions.
\newblock {\em Comptes {Rendus} {Hebdomadaires} des {Seances} de {L}'
  {Academie} des {Sciences}}, 239(15):847--848, 1954.

\bibitem{sen1924grenzbedingungen}
Nikhilranjan Sen.
\newblock {\'U}ber die grenzbedingungen des schwerefeldes an
  unstetigkeitsflächen.
\newblock {\em Annalen der Physik}, 378(5-6):365--396, 1924.

\bibitem{senovilla2013junction}
{Jos{\'e}}~{M}{M} Senovilla.
\newblock Junction conditions for f(r) gravity and their consequences.
\newblock {\em Physical Review D}, 88(6):064015, 2013.

\bibitem{senovilla2015double}
{Jos{\'e}}~{M}{M} Senovilla.
\newblock Double layers in gravity theories.
\newblock In {\em Journal of {Physics}: {Conference} {Series}}, volume 600,
  page 012004. I{O}{P} {Publishing}, 2015.

\bibitem{senovilla2018equations}
{Jos{\'e}}~{M}{M} Senovilla.
\newblock Equations for general shells.
\newblock {\em Journal of {High} {Energy} {Physics}}, 2018(11):134, 2018.

\bibitem{sfetsos1995gravitational}
{Konstadinos} Sfetsos.
\newblock On gravitational shock waves in curved spacetimes.
\newblock {\em Nuclear {Physics} {B}}, 436(3):721--745, 1995.

\bibitem{silva2024new}
JA~Silva, FC~Carvalho, and ARG Garcia.
\newblock New insights on the signature change via the colombeau framework.
\newblock {\em Classical and Quantum Gravity}, 41(19):195001, 2024.

\bibitem{steinbauer2010geometric}
R~{Steinbauer}.
\newblock A geometric approach to full {Colombeau} algebras.
\newblock {\em Banach {Center} {Publications}}, 88:267--272, 2010.

\bibitem{steinbauer2006use}
{Roland} Steinbauer and {James}~{A} {Vickers}.
\newblock The use of generalized functions and distributions in general
  relativity.
\newblock {\em Classical and Quantum Gravity}, 23(10):R91, 2006.

\bibitem{stephani2009exact}
Hans Stephani, Dietrich Kramer, Malcolm MacCallum, Cornelius Hoenselaers, and
  Eduard Herlt.
\newblock {\em Exact solutions of Einstein's field equations}.
\newblock Cambridge university press, 2009.

\end{thebibliography}

\end{document}